\documentclass[structabstract]{aa}  
\usepackage{graphicx}
\usepackage{amsmath}
\usepackage{natbib}
\bibpunct{(}{)}{;}{a}{}{,}
\usepackage{txfonts}
\begin{document}
   \title{Oscillatory dynamos and their induction mechanisms}

   \author{M. Schrinner\inst{1}\and L. Petitdemange\inst{2}\and 
   E. Dormy\inst{1}}

   \institute{MAG (ENS / IPGP), LRA, Ecole Normale Sup\'erieure, 24 rue Lhomond,
              75252 Paris Cedex 05, France\\
              \email{martin@schrinner.eu}
   \and
              Max-Planck-Institut f\"ur Astronomie, K\"onigstuhl 17, 69117 Heidelberg} 
   \date{Received; accepted}

 
  \abstract
   {Large-scale magnetic fields resulting from hydromagnetic dynamo action 
    may differ substantially in their time dependence. Cyclic field 
    variations, characteristic for the solar magnetic field, are often explained by an
    important \(\Omega\)-effect, i.e. by the stretching of field lines due to 
    strong differential rotation.}  
   {The dynamo mechanism of a convective, oscillatory dynamo model is 
    investigated.}
   {We solve the MHD-equations for a conducting Boussinesq fluid in a rotating 
    spherical shell. For a resulting oscillatory model, dynamo 
    coefficients have been computed with the help of the so-called test-field 
    method. Subsequently, these coefficients have been used in a mean-field 
    calculation in order to explore the underlying dynamo mechanism.}
   {The oscillatory dynamo model under consideration is of 
    \(\alpha^2\Omega\)-type.
    Although the rather strong differential rotation present in this model 
    influences the magnetic field, the \(\Omega\)-effect alone is not 
    responsible for its cyclic time variation. If the \(\Omega\)-effect is
    suppressed the resulting \(\alpha^2\)-dynamo remains oscillatory. 
    Surprisingly, the corresponding \(\alpha\Omega\)-dynamo leads to a 
    non-oscillatory magnetic field.}
   {The assumption of an \(\alpha\Omega\)-mechanism does not explain
    the occurrence of magnetic cycles satisfactorily.}


   \maketitle
%

\section{Introduction}
The study of the solar cycle has motivated dynamo theory for many 
decades. Hence, the solar dynamo has become the prototype of
oscillatory dynamos. However, its explanation is still 
controversial \citep{jones10}. Most solar dynamo models have been built on 
the assumption of an \(\alpha\Omega\)-dynamo mechanism \citep{ossen03}; 
that is, the poloidal field results from the interaction of helical turbulence 
with the toroidal field (\(\alpha\)-effect) whereas the toroidal field is 
thought to originate from the shearing of poloidal field lines by strong 
differential rotation (\(\Omega\)-effect). This attempt is attractive for 
mainly two reasons: 

First, the existence of a strong shear layer at the bottom of the solar 
convection zone is observationally well established, and the importance of a
resulting \(\Omega\)-effect is non-controversial. 

Second, Parker's plane layer model \citep{parker55} and in particular 
mean-field electrodynamics \citep{steenbeck66} provide a very elegant 
theoretical framework for this approach. Within mean-field theory, attention 
is focused on large scale, i.e. averaged fields, only, and the induction 
equation may be replaced by a mean-field dynamo equation \citep{krause80}
\begin{equation}
  \frac{\partial\overline{\vec{B}}}{\partial t}
   =\nabla\times(\vec{\mathcal{E}}+\overline{\vec{V}}\times\overline{\vec{B}}
   -\eta\nabla\times\overline{\vec{B}}) \, ,
  \label{eq:2}
\end{equation} 
in which \(\overline{\vec{B}}\) and \(\overline{\vec{V}}\) denote 
the average magnetic and the average velocity field, \(\eta\) stands for the 
magnetic diffusivity and \(\vec{\mathcal{E}}\) is the mean 
electromotive force. Moreover, it is assumed that 
\(\vec{\mathcal{E}}\) is homogeneous in the mean magnetic field and 
may be replaced by a parameterisation in terms of 
\(\overline{\vec{B}}\) and its first derivatives
\begin{equation}
\vec{\mathcal{E}}=\tens{a}\overline{\vec{B}}+\tens{b}\nabla\overline{\vec{B}}\,
.
\label{eq:4}
\end{equation}
In (\ref{eq:4}), the so-called dynamo coefficients \(\tens{a}\) and 
\(\tens{b}\) are tensors of second and third rank, respectively, and 
depend only on the velocity field and the magnetic diffusivity. The 
traditional \(\alpha\)-effect implemented in a large number of solar dynamo 
models \citep[e.g.][]{steenbeck66b,roberts72a,roberts72,stix76,ossen03,branden05,chan08}
corresponds to the isotropic component of \(\tens{a}\) in 
relation (\ref{eq:4}), while the \(\Omega\)-effect results from the 
\(\phi\)-component of the
\(\nabla\times(\overline{\vec{V}}\times\overline{\vec{B}})\) term in 
equation (\ref{eq:2}). 

However, a strong differential rotation is not a necessary condition for 
oscillatory solutions of the dynamo equation (\ref{eq:2}), as has been 
demonstrated in several papers \citep[see e.g.][]{raedler87,schubert00,
ruediger03,stefani03}. These authors construct models in which the toroidal 
field is likewise generated from the poloidal field by an 
\(\alpha\)-effect (\(\alpha^2\)-models) and investigate necessary constraints
on \(\tens{a}\), the boundary conditions for the magnetic field and 
the geometry of the dynamo region in order to obtain oscillatory solutions of 
(\ref{eq:2}). Recently, oscillatory dynamo models have also been investigated 
by means of direct numerical simulations. \cite{mitra10} performed dynamo 
simulations in a wedge-shaped spherical shell with an applied forcing and 
demonstrate again the existence of oscillatory \(\alpha^2\)-dynamo models. 

However, the success of mean-field models in reproducing solar-like 
variations of the magnetic field relies partly on the large number of free 
parameters, i.e. on the arbitrary determination of the dynamo coefficients 
\(\tens{a}\) and \(\tens{b}\). An alternative approach is presented by 
\cite{petrelis09}. They construct amplitude equations 
guided from symmetry considerations and analyse polarity reversals 
and oscillatons of the magnetic field resulting from the interaction  
between two dynamo modes. 

Self-consistent, global, convective 
dynamo models with cyclic magnetic field variations have
been bublished by \cite{busse06}, and \cite{goudard08}.
Convective dynamo simulations with stress-free mechanical boundary
conditions \citep{busse06} exhibit a strong and a weak field 
branch, depending on the initial conditions for the magnetic field. 
If the magnetic field is initially weak, stress free boundary conditions 
enable the development of a strong zonal flow carrying most of the kinetic 
energy and rendering convection ineffective. The magnetic field resulting from
these dynamos is rather small scaled, often of quadrupolar symmetry and weak. 
Oscillatory solutions of the induction equation are typical for this dynamo 
branch.  

A transition from steady to oscillatory dynamos may also be governed by the 
width of the convection zone; \cite{goudard08} found oscillatory models by 
decreasing the shell width. In this study, we follow their approach and 
analyse the dynamo mechanism for these oscillatory models. In particular, we 
address the question whether an \(\Omega\)-effect is responsible for the 
cyclic variation of the magnetic field. Different from previous work, we 
determine the dynamo coefficients \(\tens{a}\) and \(\tens{b}\) 
from direct numerical simulations with the help of the test-field method 
\citep{schrinner05,schrinner07}. The application of \(\tens{a}\) and 
\(\tens{b}\) in a mean-field calculation reveals their importance for 
the generation of the magnetic field.
   
\section{Dynamo calculations}
We consider a conducting Boussinesq fluid in a rotating spherical shell and  
solve the equations of magnetohydrodynamics for the velocity \(\vec{v}\), 
magnetic field \(\vec{B}\) and temperature \(T\) as given by \cite{goudard08} 
with the help of the code PaRoDy (\cite{dormy98} and further developments),
\begin{align}
E\left(\frac{\partial\vec{v}}{\partial t}+\vec{v}\cdot\nabla\vec{v}-\nabla^2\vec{v}\right)
+2\vec{z}\times\vec{v}+\nabla P =\nonumber\\
Ra\frac{\boldsymbol{r}}{r_o}T
+\frac{1}{Pm}(\nabla\times\vec{B})\times\vec{B}\, ,\\
\frac{\partial\vec{B}}{\partial t}= \nabla\times(\vec{v}\times\vec{B})
+\frac{1}{Pm}\nabla^2\vec{B} \, ,\\
\frac{\partial T}{\partial t}+(\vec{v}\cdot\nabla)(T+T_s)  =
\frac{1}{Pr}\nabla^2 T \, .
\end{align} 
Governing parameters are the Ekman number \(E=\nu/\Omega L^2\), 
the (modified) Rayleigh number \(Ra=\alpha_T g_0\Delta T L/\nu\Omega\), 
the Prandtl number \(Pr=\nu/\kappa\) and the magnetic Prandtl number 
\(Pm=\nu/\eta\). In these expressions,  \(\nu\) denotes the kinematic 
viscosity, \(\Omega\) the rotation rate, \(L\) the shell width, 
\(\alpha_T\) the thermal expansion coefficient, \(g_0\) is the gravitational 
acceleration at the outer boundary, \(\Delta T\) stands for the temperature 
difference between the spherical boundaries, \(\kappa\) is the 
thermal and \(\eta=1/\mu\sigma\) the magnetic diffusivity with the magnetic 
permeability \(\mu\) and the electrical conductivity \(\sigma\). Furthermore,
the aspect ratio is defined as the ratio of the inner to the outer shell 
radius, \(r_i/r_o\); it determines the shell width. 

In our models, convection is driven by an imposed temperature gradient 
between the inner and the outer shell boundary. The mechanical boundary 
conditions are no slip at the inner and stress free at the outer boundary. 
Moreover, the magnetic field is assumed to continue as a potential field 
outside the fluid shell. 
 
Time-averaged dynamo coefficients for an axisymmetric mean magnetic field
have been determined from direct numerical simulations as
described in detail by \cite{schrinner07} and as recently discussed for 
time-dependent dynamo models by \cite{schrinner11}. In a second step, these 
coefficients have been applied in a mean-field model based 
on equation (\ref{eq:2}) written as an eigenvalue problem,
\begin{equation}
\sigma\overline{\vec{B}}=\nabla\times\tens{D}\overline{\vec{B}} \, ,
\label{eq:5}
\end{equation}
in which the linear operator \(\tens{D}\) is defined as
\begin{equation}
\tens{D}\overline{\vec{B}}=\overline{\vec{V}}\times\overline{\vec{B}}+\tens{a}\overline{\vec{B}}+\tens{b}\nabla\overline{\vec{B}}-\frac{1}{Pm}\nabla\times\overline{\vec{B}}
\, .
\label{eq8}
\end{equation} 
The time evolution of each mode is determined by its eigenvalue \(\sigma\) and 
proportional to \(\exp{(\sigma t)}\). For more details concerning the 
eigenvalue calculation, we refer to \cite{schrinner10b}. 

We also consider the evolution of a kinematically advanced magnetic field, 
\(\vec{B}_\mathrm{Tr}\), governed by a second induction equation  
\begin{equation}
\frac{\partial\vec{B}_\mathrm{Tr}}{\partial
  t}=\nabla\times(\vec{v}\times\vec{B}_\mathrm{Tr})+\frac{1}{Pm}\nabla^2\vec{B}_\mathrm{Tr}\,
.
\label{eq:6}
\end{equation}
The tracer field \(\vec{B}_\mathrm{Tr}\) experiences the self-consistent
velocity field at each time step but does not contribute to the Lorentz force 
and is passive in this sense \citep[see also][]{schrinner10a}. Its evolution 
will be compared with mean-field results originating likewise 
from a kinematic approach. Moreover, a kinematically advanced
tracer field allows us to test for the influence of the \(\Omega-\)effect in
direct numerical simulations. In a numerical experiment, we subtract the
contribution of the \(\Omega-\)effect and the mean meridional flow in the equation for the tracer field,
\begin{equation}
\frac{\partial\vec{B}_\mathrm{Tr}}{\partial
  t}=\nabla\times(\vec{v}\times\vec{B}_\mathrm{Tr})+\frac{1}{Pm}\nabla^2\vec{B}_\mathrm{Tr}-\nabla\times(\overline{\vec{V}}\times\overline{\vec{B}}_\mathrm{Tr})
\, ,
\label{eq:8}
\end{equation} 
and study in this way the outcome of a kinematic \(\alpha^2\)-dynamo.
\section{Results}
The model under consideration has been previously studied by \cite{goudard08}. It is defined
by \(E=10^{-3}\), \(Ra=100\,(=2.8\,Ra_c)\), \(Pm=5\), \(Pr=1\) and an aspect 
ratio of \(0.65\). Except for the stress-free mechanical boundary condition 
applied at \(r=r_o\) and an increased aspect ratio, the governing parameters 
are those of a rather simple, quasi-steady benchmark dynamo 
\citep{christensen01}. However, \cite{goudard08} report a transition from 
steady, dipolar to oscillatory models for these parameter values. Note that
the model requires a rather high angular resolution up to harmonic degree 
\(l_\mathrm{max}=112\).

\begin{figure}
  \centering
  \includegraphics{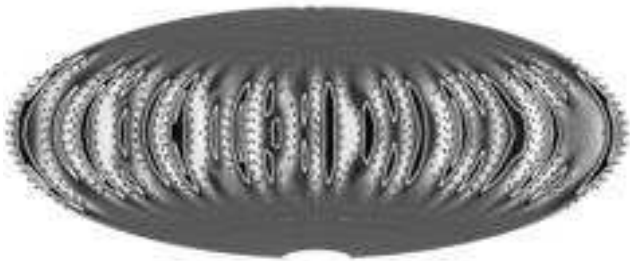}
  \caption{Snapshot of the radial velocity of the considered dynamo model at
           \(r=0.79\,r_o\). The velocity component has been normalised by its 
           maximum absolute value, \(\vec{v}_{r,\,\mathrm{max}}=24.46\,\nu/L\). Hence, 
           the colour coding ranges from \(-1\), white, to \(+1\), black. 
           Contour lines correspond to \(\pm 0.2\) and \(\pm 0.6\).  
  }
  \label{fig1}
\end{figure}

Figure \ref{fig1} displays the radial component of the velocity field at a 
given radial level. A typical columnar convection pattern is visible, even 
though the convection columns are noticeably disturbed by the influence 
of the curved boundaries and a strong zonal flow carrying about
50 per cent of the kinetic energy. The magnetic Reynolds number based on the 
rms-velocity and the shell width, \(Rm=v_{\mathrm{rms}}\,L/\eta\), is about 90.
The flow is symmetric with respect to the equatorial plane and convection 
takes place only outside the inner core tangent cylinder.    

\begin{figure}
  \centering
  \includegraphics{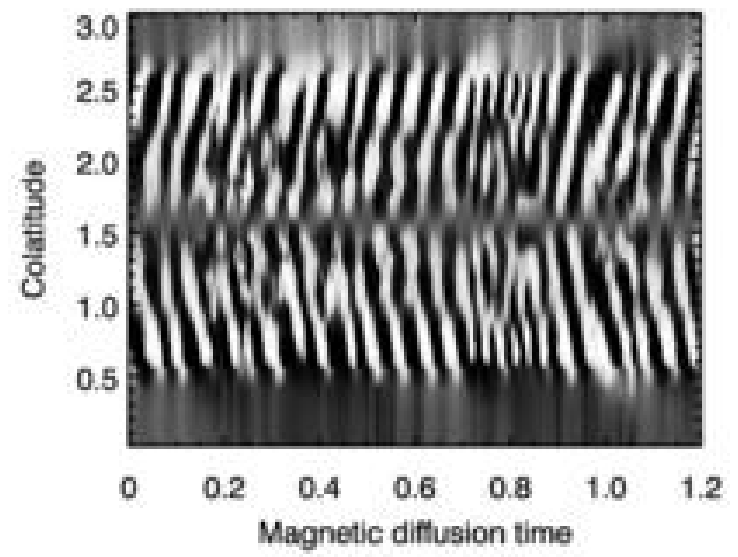}
  \includegraphics{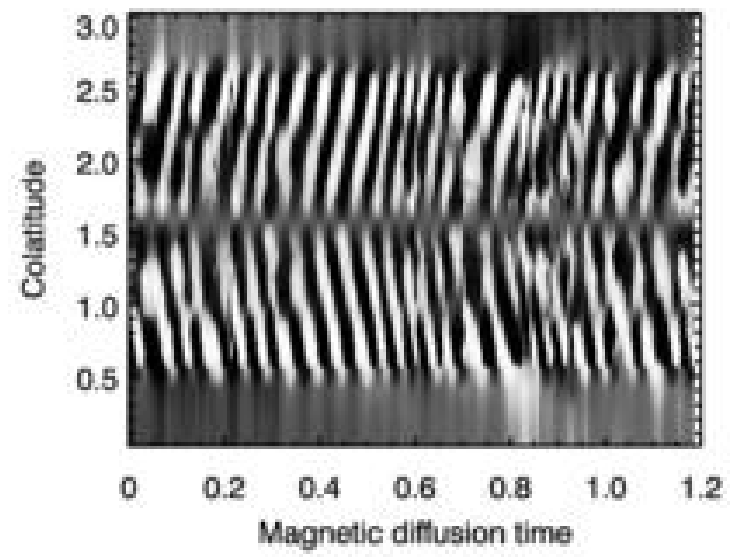}
  \includegraphics{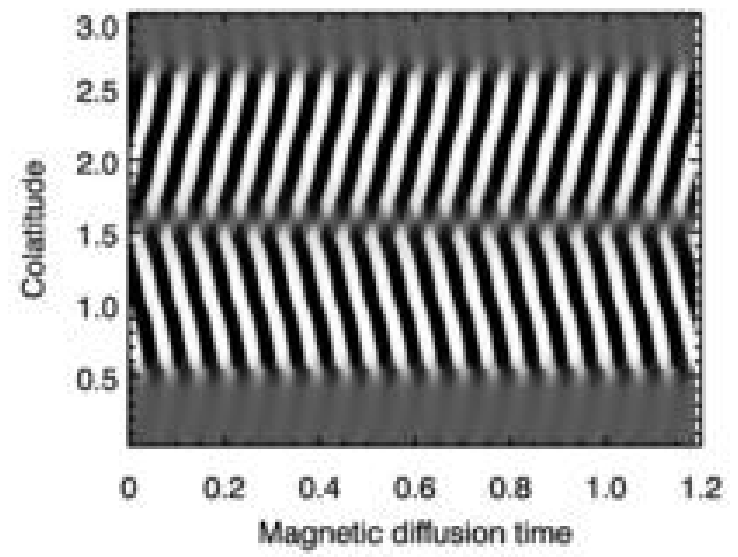}
  \caption{Azimuthally averaged radial magnetic field at the outer shell 
  boundary varying with time (butterfly diagram) resulting from a 
  self-consistent calculation (top), kinematic calculation according to 
  (\ref{eq:6}) (middle) and 
  mean-field calculation (bottom). The contour plots have been normalised by 
  their maximum absolute value at each time step considered. The colour coding 
  ranges from \(-1\), white, to \(+1\), black.  
  }
  \label{fig2}
\end{figure} 
The evolution of the magnetic field is cyclic. In figure \ref{fig2}
(top), contours of the azimuthally averaged radial magnetic field at the outer 
shell boundary varying with time are plotted in a so-called butterfly diagram.
A dynamo wave migrates away from the equator until it reaches 
mid-latitudes where the inner core tangent cylinder intersects the outer shell 
boundary. The magnetic field looks rather small scaled and multipolar.
This is confirmed by the magnetic energy spectrum which is essentially white, 
except for a negligible dipole contribution. Furthermore, the magnetic field is
weak, as expressed by an Elsasser number of 
\(\Lambda=B^2_{\mathrm{rms}}/(\mu\rho\eta\Omega)=0.13\). 

The kinematically advanced tracer field grows slowly in time, i.e. the model 
under consideration is kinematically unstable according to the classification
by \cite{schrinner10a}. But, deviations of the tracer field from the actual
field are hardly noticeable in the field morphology. Moreover, the very same
dynamo wave persists in the kinematic calculation (see also \cite{goudard08}), as visible in figure
\ref{fig2} (middle). Note that the tracer field in figure \ref{fig2} has 
evolved from random initial conditions. 

A mean-field calculation based on the dynamo coefficients 
\(\tens{a},\,\tens{b}\) and the mean flow \(\overline{\vec{V}}\) determined 
from the self-consistent model is presented in the bottom line of figure 
\ref{fig2}. The fastest growing eigenmodes form a conjugate complex pair and 
give rise to a dynamo wave which compares nicely with the direct numerical 
simulations. Since this model depends on the full \(\tens{a}\)-tensor 
and the mean flow, we refer to it as an \(\alpha^2\Omega\)-dynamo. 

\begin{figure}
  \centering
  \includegraphics{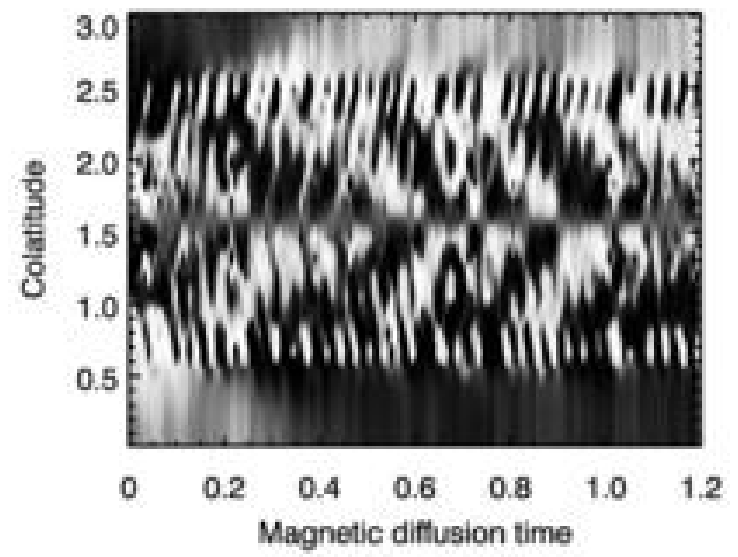}
  \includegraphics{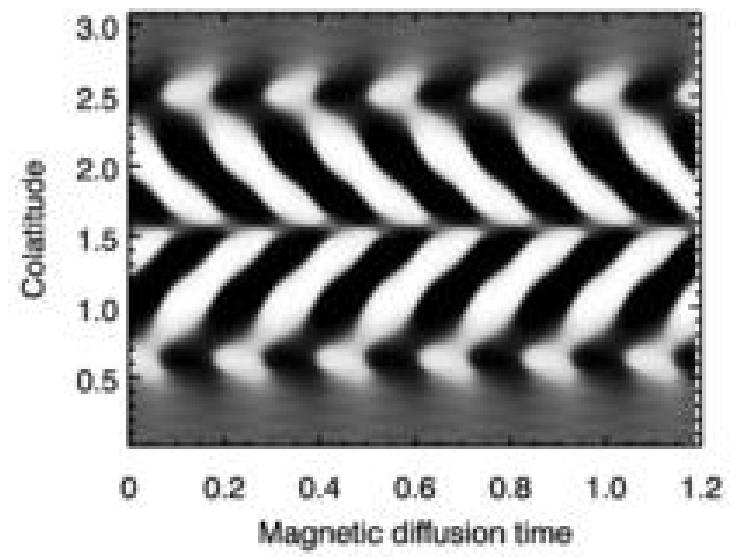}
  \caption{Azimuthally averaged radial magnetic field at the outer shell 
  boundary varying with time (butterfly diagram) resulting from a 
  kinematic calculation with subtracted \(\Omega\)-effect (top) and 
  a corresponding mean-field calculation (bottom). The contour plots are 
  presented as in figure (\ref{fig2}).  
  }
  \label{fig3}
\end{figure} 

The influence of the differential rotation may be suppressed in the kinematic 
calculation of the tracer field without changing any other component of the flow. A 
butterfly diagram resulting from a kinematically advanced field according 
to equation (\ref{eq:8}) is presented in figure \ref{fig3} (top). The evolution
of the magnetic field is again cyclic. Apart from small-scale variations on  
shorter time scales, a dynamo wave migrates from mid-latitudes towards 
the equator. This is in agreement with a corresponding mean-field calculation in 
which the mean flow \(\overline{\vec{V}}\) in (\ref{eq8}) has been canceled: 
The bottom chart of figure \ref{fig3} provides the butterfly diagram stemming from the fastest 
growing eigenmodes of the resulting \(\alpha^2\)-dynamo. An explanation why direct 
numerical simulations and mean field calculations compare somewhat better in figure 
\ref{fig2} than in figure \ref{fig3} is provided in appendix \ref{appendix2}.     

\begin{figure}[t]
  \centering
  \includegraphics{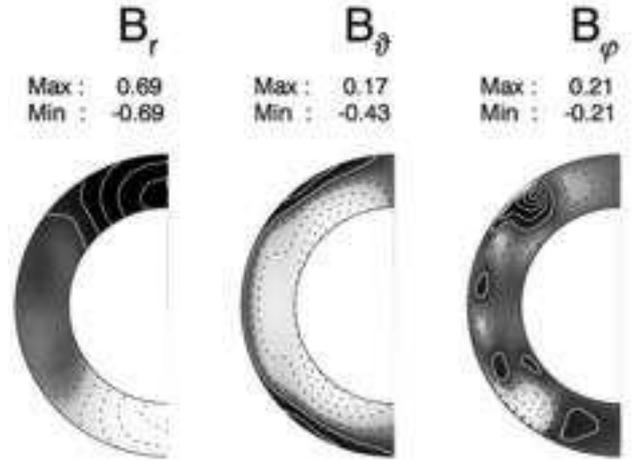}
  \caption{The leading dipolar eigenmode resulting from a mean-field calculation with 
           \(a_{rr}=a_{\theta\theta}=0\). Contour plots of all three
           components are presented, each 
           normalised separately by their maximum absolute values. Maxima and minima
           are written next to each plot. The colour coding ranges from \(-1\), white, to \(+1\), black,
           and contour lines correspond to \(\pm 0.1\), \(\pm 0.3\), \(\pm 0.5\), \(\pm 0.7\) and \(\pm 0.9\).    
   }
  \label{fig4}
\end{figure}

The time evolution of the related \(\alpha\Omega\)-dynamo is of further interest.
As the \(\alpha\)-effect is not directly accessible in direct numerical simulations,
the corresponding \(\alpha\Omega\)-dynamo can be realised in a mean-field calculation, only.
In a first attempt, we have set \(\tens{a}_{rr}=\tens{a}_{\theta\theta}=0\) in order to suppress the generation 
of toroidal field from poloidal field by an \(\alpha\)-effect. Both components make major 
contributions to this process. The leading eigenmode resulting from this calculation is shown in 
figure \ref{fig4}; it is real, i.e. non-oscillatory, and close to
marginal stability. The results remain similar if we neglect further, non-diagonal components of \(\tens{a}\).

\section{Discussion}
The Frequency and the propagation direction of the dynamo wave visible in figure \ref{fig2} depend 
strongly on  the differential rotation, in agreement with \cite{busse06}. We 
follow their approach and give an estimate for the cycle frequency applying Parker's plane layer 
formalism \citep{parker55}. To this end, we introduce a Cartesian coordinate system \((x,y,z)\) 
corresponding to the \((\phi,\theta,r)\) directions and define mean quantities to be x-independent. 
Moreover, we write 
\(\overline{\vec{B}}=B\vec{e_x}+\vec{B_p}=B\vec{e_x}+\nabla\times A\vec{e_x}\) and reformulate 
(\ref{eq:2}) in the following simplified manner
\begin{eqnarray}
\frac{\partial A}{\partial
  t}&=&\tens{\alpha}_{xx}B+\frac{1}{Pm}\nabla^2A\, ,\label{eq:10}\\
\frac{\partial B}{\partial t}&=&-\frac{\partial}{\partial
  y}(\tens{\alpha}_{zz}\frac{\partial A}{\partial
  y})+\frac{\mathrm{d}\overline{\vec{V}}_x}{\mathrm{d}z}\frac{\partial
  A}{\partial y}+\frac{1}{Pm}\nabla^2B \, .
\label{eq:12}
\end{eqnarray}  
In the above equations, we have considered only the dominant diagonal components of \(\tens{a}\), 
\(\tens{a}_{rr}\) and \(\tens{a}_{\phi\phi}\) corresponding to \(\tens{\alpha}_{zz}\) and 
\(\tens{\alpha}_{xx}\); all components of \(\tens{b}\) and the mean meridional flow have been 
neglected. Furthermore, \(\overline{\vec{V}}\) has been assumed to depend only on \(z\). Then, the 
ansatz
\begin{equation}
(A,B)=(\hat{A},\hat{B})\exp(i\vec{k}\cdot \vec{x}+\sigma t) \, ,
\label{eq:13}
\end{equation}
leads to 
\begin{eqnarray}
p\hat{A} & = & \alpha_{xx}\hat{B} \, ,\label{eq:26}\\
p\hat{B} & = &
\left(\alpha_{zz}k_y^2+ik_y\left(\frac{\mathrm{d}\overline{\vec{V}}}{\mathrm{d}z}-\frac{\partial\alpha_{zz}}{\partial
  y}\right)\right)\hat{A}\, ,\label{eq:28}
\end{eqnarray}
with \(p=\sigma+|\vec{k}|^2/Pm\). From (\ref{eq:26}), (\ref{eq:28}), we derive a dispersion relation 
\begin{equation}
p^2=\alpha_{xx}\left(\alpha_{zz}k_y^2+ik_y\left(\frac{\mathrm{d}\overline{\vec{V}}}{\mathrm{d}z}-\frac{\partial\alpha_{zz}}{\partial
  y}\right)\right) \, ,
\label{eq:30}
\end{equation}  
from which the real and the imaginary part of \(\sigma\) can be calculated. If \(\alpha_{xx}\) 
is  positive (e.g. in the northern hemisphere),  it follows 
\begin{eqnarray}
\lambda&=&\Re(\sigma)=-|\vec{k}|^2/Pm\nonumber\\
&&+\sqrt{\frac{\alpha_{xx}}{2}}\left(\sqrt{(\alpha_{zz}k_y^2)^2+\left(\frac{\mathrm{d}\overline{\vec{V}}}{\mathrm{d}z}-\frac{\partial\alpha_{zz}}{\partial
    y}\right)^2 k_y^2}+\alpha_{zz}k_y^2\right)^{1/2} \, 
\end{eqnarray}
and
\begin{eqnarray}
\omega&=&\Im(\sigma)=\nonumber\\
&&\pm\sqrt{\frac{\alpha_{xx}}{2}}\left(\sqrt{(\alpha_{zz}k_y^2)^2+\left(\frac{\mathrm{d}\overline{\vec{V}}}{\mathrm{d}z}-\frac{\partial\alpha_{zz}}{\partial
    y}\right)^2 k_y^2}-\alpha_{zz}k_y^2\right)^{1/2} \, .
\label{eq:32}
\end{eqnarray}
The sign in (\ref{eq:32}) is determined by the sign of \(k_y(\mathrm{d}\overline{\vec{V}}/\mathrm{d}z-\partial\alpha_{zz}/\partial y)\). If we further assume that the frequency is dominated by differential 
rotation and neglect the \(\alpha\)-terms in (\ref{eq:32}), we estimate similar to \cite{busse06}
\begin{equation}
\omega=\Im(\sigma)\approx\pm\left(\frac{\pi}{L^2}\tens{\alpha}_{xx}\sqrt{2\overline{E}_T}\right)^{1/2}
\, .
\label{eq:14}
\end{equation} 
In (\ref{eq:14}), \(\overline{E}_T\) denotes the the kinetic energy density due to the axisymmetric 
toroidal velocity field and \(k_y\approx 2\pi/L\) has been used. 
Approximating \(\tens{\alpha}_{xx}\) by the rms-value of 
\(\tens{a}_{\phi\phi}\), \(\tens{\alpha}_{xx}=11.50\,\nu/L\), and with  \(E_T=71\,\nu^2/L^2\), we
find \(\omega\approx \pm 103.7\,\eta/L^2\) which is surprisingly close to 
\(\omega=\pm 100.95 \eta/L^2\) in the full calculation presented in figure \ref{fig2}. Note that 
\(\tens{\alpha}_{xx}\) and \(\mathrm{d}\overline{\vec{V}}/\mathrm{d}z\) are of the same order of magnitude
and contribute equally to \(\omega\). The sign in (\ref{eq:14}) is determined by the sign of the 
product \(\tens{a}_{\phi\phi}\,\partial\overline{\vec{V}}_\phi/\partial r\) which is positive in 
the northern and negative in the southern hemisphere. Therefore, our estimate in (\ref{eq:14}) predicts 
a dynamo wave migrating away from the equator. This is in agreement with the simulations shown in 
figure \ref{fig2}. 

\begin{figure}
  \centering
  \includegraphics{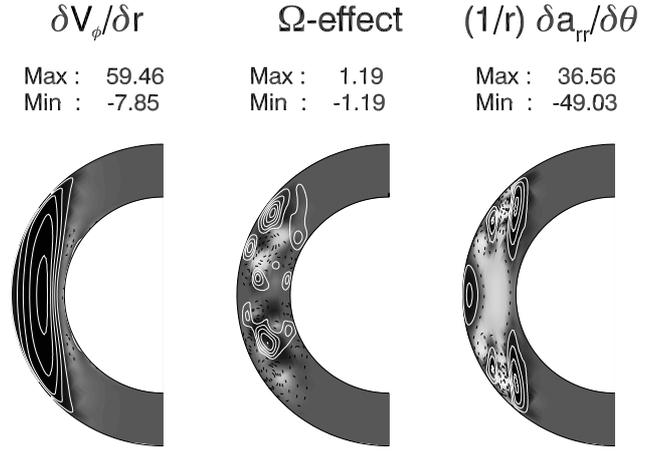}
  \caption{
  Left: \(\partial\overline{\vec{V}}_\phi/\partial r\) 
  in units of \(\nu/L^2\). Middle: \(\Omega\)-effect as given by
  \(r\overline{\vec{B}}_r\,\partial(r^{-1}\overline{\vec{V}}_\phi)/\partial r+r^{-1}\sin\theta\,\overline{\vec{B}}_\theta\,\partial(\sin\theta^{-1}\overline{\vec{V}}_\phi)/\partial\theta\) (greyscale) and 
  \(\overline{\vec{B}}_\phi\) (superimposed contour lines, solid [dashed] lines indicate positive 
  [negative] values). 
  Right: \((1/r)\,\partial\tens{a}_{rr}/\partial\theta\) in units of \(\nu/L^2\).  
  The contour plots are presented in the same style as in figure \ref{fig4}. 
          }
  \label{fig5}
\end{figure}

However, the attempt to describe the model under consideration as an \(\alpha\Omega\)-dynamo
fails. An oscillatory mode with a frequency close to the above estimate turns out to be 
clearly subcritical in a mean-field calcuation, if \(\tens{a}_{rr}\) 
and \(\tens{a}_{\theta\theta}\) are omitted. Instead, this model is governed by a real, 
dipolar mode close to marginal stability (see figure \ref{fig4}). Hence, the \(\Omega\)-effect 
is only partly responsible for the generation of the mean azimuthal field, as confirmed by 
figure \ref{fig5}. The chart in the middle compares the 
\(\Omega\)-effect, \(r\overline{\vec{B}}_r\,\partial(r^{-1}\overline{\vec{V}}_\phi)/\partial r+r^{-1}\sin\theta\,\overline{\vec{B}}_\theta\,\partial(\sin\theta^{-1}\overline{\vec{V}}_\phi)/\partial\theta\) in greyscale with the mean azimuthal field displayed by superimposed contour lines. In   
particular, the elongated flux patches close to the inner core tangent cylinder are, if at all, 
negatively correlated with the \(\Omega\)-effect. Consistent with this finding, the poloidal 
axisymmetric magnetic energy density exceeds the toroidal one by 20\%. 

Differential rotation alone is not responsible for the cyclic time evolution of the magnetic 
field, despite its influence on the frequency and the propagation direction of the dynamo wave.
This is most clearly visible in figure \ref{fig3}. Simulations without differential rotation 
still lead to a dynamo wave even though its frequency and propagation direction have changed. 
In the framework of Parker's plane layer formalism, the frequency of this oscillatory 
\(\alpha^2\)-dynamo depends crucially on \(-\partial\tens{\alpha}_{zz}/\partial y\) instead
of \(d\overline{\vec{V}}_x/dz\).
Note the additional minus sign, which might explain the reversed propagation direction if 
the assume that \(\partial\alpha_{zz}/\partial y\) is predominantly positive. But 
different from the radial derivative of the mean azimuthal flow, 
\((1/r)\,\partial\tens{a}_{rr}/\partial\theta\) is highly structured, changes sign in radial direction 
and exhibits localised patches of low negative values (see figure \ref{fig5}). Therefore, we do not 
attempt to give an estimate for the frequency similar to (\ref{eq:14}). 

\begin{figure}
  \centering
  \includegraphics{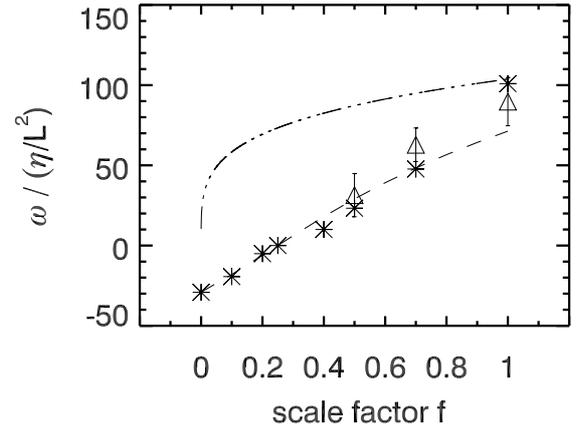}
  \caption{Frequencies resulting from kinematic calculations in which the amplitude of the mean flow 
           has been changed by multiplication with a scale factor, \(f\). Stars denote frequencies 
           stemming from an eigenvalue calculation according to (\ref{eq:5}), whereas triangles are 
           estimates obtained from kinematic results due to (\ref{eq:6}). The dashed-dotted line gives 
           frequencies as predicted for an \(\alpha\omega\)-dynamo by (\ref{eq:14}), 
           while the dashed line represents \(\omega\) as a function of \(\overline{\vec{V}}\) for an 
           \(\alpha^2\omega\)-dynamo according to (\ref{eq:32}).
           }
  \label{fig6}
\end{figure}

In order to better understand the influence of the mean flow on the frequency of the dynamo wave, we 
have gradually changed the amplitude of \(\overline{\vec{V}}\) in a series of kinematic calculations. 
Results are presented in figure \ref{fig6}. Stars denote frequencies obtained from eigenvalue 
calculations according to (\ref{eq:5}), wheras triangles stand for frequencies estimated from kinematic 
results due to equation (\ref{eq:6}). In both cases, the amplitude of the mean 
flow \(\overline{\vec{V}}\) has been varied by multiplication with a scale factor \(f\). 
For \(f=1\), the original calculation is retained, while for \(f=0\), we reproduce 
the \(\alpha^2\)-dynamo already discussed above. Frequencies of dynamo waves resulting from direct 
numerical simulations according to (\ref{eq:6}) have been meassured for \(f=1,\,0.7\), and \(0.5\). 
Owing to the turbulence present in the simulations, these are rather 
rough estimates and error bars have been included. Nevertheless, the results obtained are in satisfactory 
agreement with the eigenvalue calculations. The frequencies in figure \ref{fig6} decrease 
continously with decreasing scale factors. If the amplitude of \(\overline{\vec{V}}\) is reduced to 25 
per cent of its original value, \(\omega\) changes sign and the propagation direction of the 
dynamo wave is reversed. The dashed-dotted line in figure \ref{fig6} gives \(\omega\) according to 
(\ref{eq:14}) as predicted for an \(\alpha\omega\)-dynamo. It matches the numerical results 
if \(\mathrm{d}\overline{\vec{V}}/\mathrm{d}z\) dominates in (\ref{eq:32}) but deviates clearly for 
smaller amplitudes. On the other hand, it is illustrative to use relation (\ref{eq:32}) to model the 
dependence of \(\omega\) on the mean flow. If we set \(\partial\alpha_{zz}/dy=0.25\,\mathrm{d}\overline{\vec{V}}/\mathrm{d}z\) and determine a representative value for \(\alpha_{zz}\) inverting (\ref{eq:32}) for 
\(\mathrm{d}\overline{\vec{V}}/\mathrm{d}z=0\) and \(\omega=-29.15 \eta/L^2\), the dashed line in 
figure \ref{fig6} results from (\ref{eq:32}). It fits the numerical data rather well and converges 
towards the frequencies predicted for an \(\alpha\omega\)-dynamo, if the amplitude of 
\(\overline{\vec{V}}\) is sufficiently high.

Let us stress again that some caution is needed in applying the present mean-field analysis 
to non-linear direct numerical simulations, as the the dynamo model considered here is
kinematically unstable. Strictly speaking, our mean-field results are only relevant for the 
kinematically advanced tracer field. But, because the model is close to dynamo onset and only 
weakly non-linear, we believe that our interpretation is also valid for the fully 
self-consistent field. This is in particular confirmed by the rather good agreement of the 
three butterfly diagrams presented in figure~\ref{fig2}.

\section{Conclusions}
A particular dynamo mechanism does not seem to be responsible for the occurrence of 
periodically time-dependent magnetic fields. 
It turns out, that the influence of the large-scale radial 
shear (the \(\Omega\)-effect), is not necessary for cyclic field variations.
Instead, the action of small-scale convection, represented by a spatially structured dynamo coefficient
\(\tens{a}_{rr}\), happens to be essential.
For the model presented here, small convective 
length scales are forced by a thin convection zone. 
Further investigations are needed to assess whether our finding is representative 
for a wider class of oscillatory models.
\begin{acknowledgements}
MS is grateful for financial support from the ANR Magnet project. The computations have 
been carried out at the French national computing center CINES.
\end{acknowledgements}
\appendix
\section{The use of time averaged dynamo coefficients}
\label{appendix2}
In the following, azimuthal averages are, as throughout in the paper, denoted by an 
overbar, time averages are expressed by brackets, \(<\cdots>\). Initially, dynamo 
coefficients have been determined for an azimuthally averaged, mean magnetic 
field \(\overline{\vec{B}}\). Hence, the evolution of the latter is given by 
\begin{equation}
\frac{\partial\overline{\vec{B}}}{\partial t}
   =\nabla\times(\tens{a}\overline{\vec{B}}+\tens{b}\nabla\overline{\vec{B}}+\overline{\vec{V}}\times\overline{\vec{B}}
   -\frac{1}{Pm}\nabla\times\overline{\vec{B}}) \, .
\label{eq:16}
\end{equation} 
But, the dynamo coefficients \(\tens{a}\), \(\tens{b}\) and the mean 
flow \(\overline{\vec{V}}\) vary stochastically in time. In order to describe the average 
dynamo action, we take in addition the time average of these quantities and write 
approximatively,
\begin{equation}
\frac{\partial\overline{\vec{B}}}{\partial t}
   \approx\nabla\times(<\!\tens{a}\!>\overline{\vec{B}}+<\!\tens{b}\!>\nabla\overline{\vec{B}}+<\!\overline{\vec{V}}\!>\times\overline{\vec{B}}
   -\frac{1}{Pm}\nabla\times\overline{\vec{B}}).
\label{eq:18}
\end{equation} 
We emphasise that there is no a priori relation between the left hand side and the right 
hand side of equation (\ref{eq:18}). The actual, azimuthally averaged magnetic field will 
deviate from our mean-field description the stronger, the more \(\tens{a}\), \(\tens{b}\) 
and \(\overline{\vec{V}}\) fluctuate in time. Among these three quantities, the mean flow
\(\overline{\vec{V}}\) is almost time independent, whereas \(\tens{a}\) and \(\tens{b}\)
vary considerably. This is the reason, why the butterfly diagrams in figure 
\ref{fig2} are in better agreement than in figure \ref{fig3}, for which the stabilizing 
influence of the mean flow has been omitted.      

\bibliographystyle{aa}
\bibliography{schrinner}

\end{document}